\documentstyle[aps,multicol,graphicx]{revtex}
\begin{document}
%TCIMACRO{\TeXButton{begmulti}{\begin{multicols}{2}}}
%BeginExpansion
\begin{multicols}{2}
%EndExpansion

{\bf Comment on ``Vortex Glass and Lattice Melting Transitions in a YNi}$_2$%
{\bf B}$_2${\bf C Single Crystal''}

In a recent Letter~\cite{MOMun96a}, Mun {\em et al.} claim to observe vortex
glass and lattice melting transitions in superconducting YNi${}_2$B${}_2$C.
A key observation, which they associate with vortex lattice melting, is the
presence of a sharp transition in the dc resistivity in finite dc magnetic
field. Mun {\em et al.} arrive at an $HT$ phase diagram with a single phase
boundary which separates the vortex liquid from the vortex solid/glass and a
single $H_{c2}(T)$ Superconductor-Normal (SN) boundary. In this Comment, we
show that precision rf penetration depth measurements indicate a much more
complex picture than that presented in Ref.~\cite{MOMun96a}.

\begin{figure}[tbph]
\narrowtext
\begin{center}
  \includegraphics*[width=0.45\textwidth]{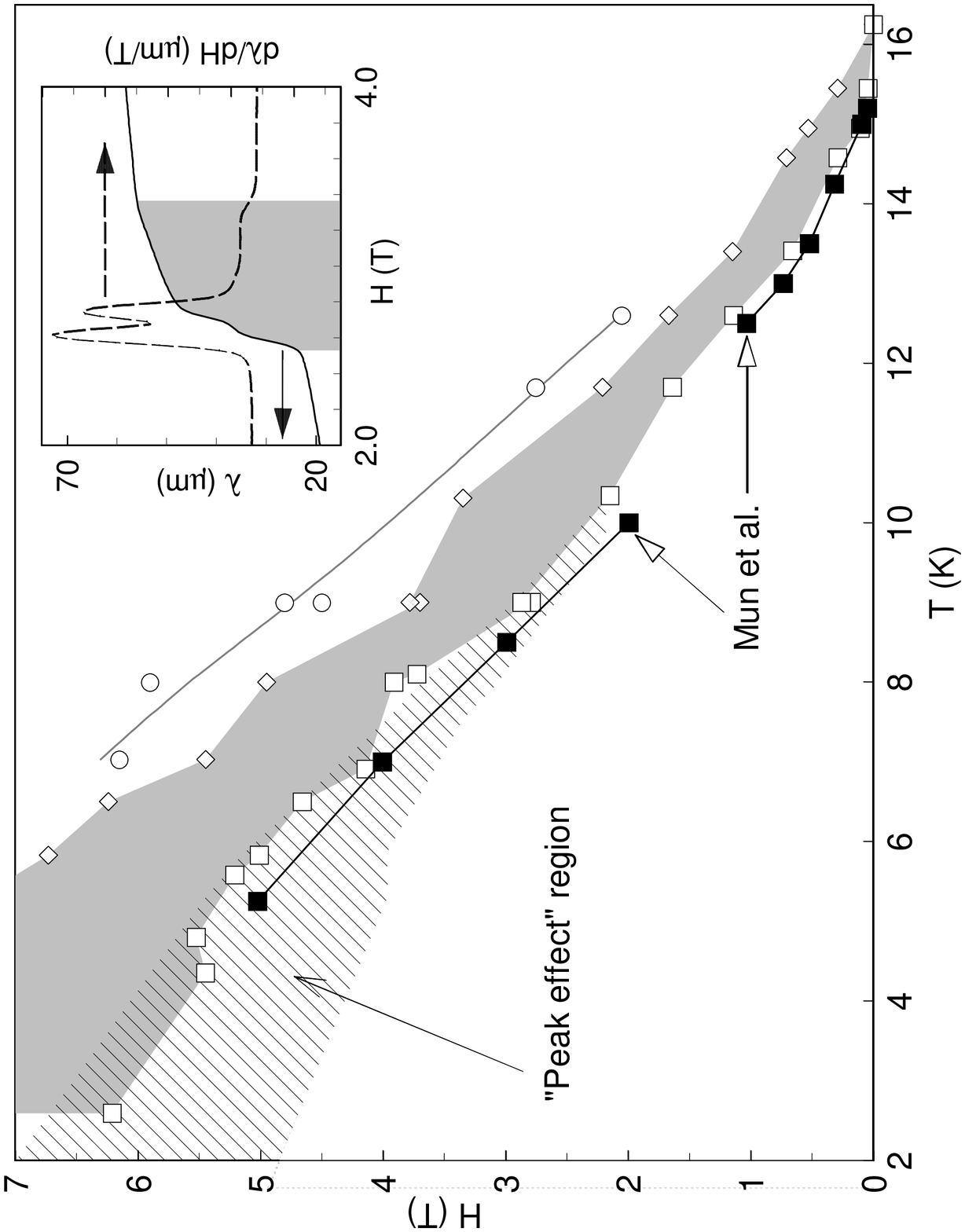} 
\end{center}
  
  \caption{
    $HT$ phase diagram for YNBC (as described in the text) from rf
    measurements.
%Two regions are
%    highligted in the phase diagram: a region where the ``peak
%    effect'' is observed in the vortex response (hatched) and a region
%    where we observe structure in the SN transition (shaded). This
%    structure often extends to the uppermost field scale (open
%    circles), although a sensitivity to surface conditioning may
%    indicate that this field is associated with surface
%    superconductivity.  
    The data from Ref.~\protect\cite{MOMun96a} is also presented
    (filled squares).  Inset: Typical $\lambda$ {\it vs.}\/ $H$ for
    $T < T{_c}$. The derivative ($d\lambda/dH$) is also shown to
    highlight the structure in the transition. The shaded band
    corresponds to the shaded region in the phase diagram.}
  \label{fig:onefig}
\end{figure}

Our experiments are carried out on samples obtained from the {\em same
source} as in Ref.~\cite{MOMun96a}, using an rf (3 MHz) tunnel diode
technique. Details of the measurement technique and particularly results on
single crystal YNi${}_2$B${}_2$C are described elsewhere~\cite{SOxx96a}.
Typical $\lambda (H)\,vs.\,H$ data observed for $T<T_c$ are shown in the
inset of Fig.\ref{fig:onefig}. The data clearly show the presence of
multiple transitions at characteristic fields and these are accentuated in
the $d\lambda /dH$ also shown in the plot. We have studied these features
over a wide range of temperatures and fields in the superconducting state.

The $HT$ phase diagram based on our results is shown in the main panel
of Fig.~\ref{fig:onefig} with $H_g(T)$ and $H_m(T)$ from
Ref.~\cite{MOMun96a} also included for comparison (filled
squares/solid line). It is obvious that the phase diagram is quite
complex~\cite {SOxx96a}, consisting of $3$, possibly $4$, transition
lines. The vortex-glass transition identified by Ref.~\cite{MOMun96a}
lies in a region of the phase diagram where we have observed a
signature of the ``peak effect'' indicative of a softening of the
vortex lattice (hatched region)\cite{SOxx96a}.
%However, the
%interesting multiple structure as depicted in the inset of Fig.~\ref
%{fig:onefig} occurs in the region (shown shaded) at fields greater
%than the melting line of Ref.~\cite{MOMun96a}. 
However, the melting line reported in Ref.~\cite{MOMun96a} appears to
coincide with the onset of a region (shown shaded) where we observe multiple
structure in the transition [Fig.~\ref{fig:onefig}(inset)], which
persists even above the glass-liquid transition reported in
Ref.~\cite{MOMun96a}. It is likely that these features have been
missed by Ref.~\cite{MOMun96a} owing to the experimental limitations
of dc measurements. It is clear that a simple interpretation in terms
of a {\em single} melting transition below a {\em single} $H_{c2}$
boundary is inadequate.

We note other important features of our data:

\begin{enumerate}
\item The multiple transitions appear to extend to the $H=0$ line
  which raises the possibility that these transitions, including the
  so-called melting line of Ref.~\cite{MOMun96a}, may be associated
  with the {\em condensate} rather than with vortices.

\item Unlike dc measurements which are shorted out by the first
  superconducting percolative path, our rf measurements probe well
  ($O(\mu {\rm m})$) into the bulk of the sample. Thus, the multiple
  features we report may best be observed in ac measurements.
  
\item Similar structure is observed in ErNi${}_2$B${}_2$C and
  HoNi${}_2$B${}_2$C also. Thus these features are common to other
  members of the borocarbide family, and apparently independent of the
  magnetic nature of the rare earth element.
  
\item Polishing the crystal surface tends to suppress the uppermost
  field scale (marked by open cicles and a solid line in
  Fig.~\ref{fig:onefig}) indicative of a certain sensitivity to the
  microstructure. However, the multiple features in the shaded region
  remain unaffected. The uppermost field scale may be associated with
  surface superconductivity.
\end{enumerate}

In conclusion, we have presented results that indicate that
YNi${}_2$B${}_2$ C, as well as other members of the borocarbide
superconductor family, possess unusual structure indicative of
multiple transitions near the SN phase boundary. The melting line of
Ref.~\cite{MOMun96a} appears to be only one of these transitions. The
origin of these transitions is not known at present. These results
appear to preclude a simple description in terms of melting of the
vortex lattice as arrived at in Ref.~\cite{MOMun96a}.

This work was supported by NSF-DMR-9623720. We thank Paul Canfield for
providing the samples and useful discussions.

\noindent 
S.~Sridhar, S.~Oxx, Balam~A. Willemsen, H.~Srikanth and D.~P. Choudhury

Physics Department, Northeastern University, \\
360 Huntington Ave., Boston, MA 02115

\today

PACS: 74.60.Ec, 74.25.Fy, 74.60.Ge, 74.70.Ad

%TCIMACRO{\TeXButton{endmulti}{\end{multicols}}}
%BeginExpansion 
\end{multicols}
%EndExpansion
\end{document}